%% file: pqca.tex
\documentclass{llncs}

\input{braket}

\usepackage{graphicx}
\usepackage{amsfonts}
\usepackage{amsmath}
\usepackage{latexsym}
\usepackage{amssymb}
\usepackage[mathcal]{euscript}
\usepackage[table]{xcolor}

\newcommand\VL[1]{#1} 	
\newcommand\VS[1]{} 	
\newcommand\VLtodo[1]{}	
\newcommand\couic[1]{}

\newcommand{\Z}{\mathbb{Z}}

\newcommand{\joliH}{\mathcal{H}}

\newcommand{\joliN}{\mathcal{N}}
\newcommand{\pa}[1]{\left(#1\right)}
\newcommand{\acco}[1]{\left\{#1\right\}}

\newcommand{\etal}{\emph{et al\ }{}}
\newcommand{\eg}{\emph{e.g.\ }{}}
\newcommand{\ie}{\emph{i.e.\ }{}}
\newcommand{\cf}{\emph{cf.\ }{}} 
\newcommand{\videinfra}{\emph{vide infra}{}} 

\newtheorem{Th}{Theorem}
\newtheorem{Cl}{Claim}

\newtheorem{Def}{Definition}

\begin{document}

\title{
Partitioned quantum cellular automata\\ are intrinsically universal
}

\author{Pablo Arrighi \and Jonathan Grattage}

\institute{University of Grenoble, \VL{Laboratoire }LIG,\VL{\\}
 \VL{B\^{a}timent IMAG C, }220 rue de la Chimie,\VL{\\}
 38400 \VS{SMH}\VL{Saint-Martin-d'H\`eres}, France
 \and
 Ecole Normale Sup\'erieure de Lyon, \VL{Laboratoire }LIP,\VL{\\}
 46 all\'ee d'Italie, 69364 Lyon cedex 07, France}

\maketitle
 
\begin{abstract}
There have been several non-axiomatic approaches taken to define Quantum Cellular Automata (QCA). Partitioned QCA (PQCA) are the most canonical of these non-axiomatic definitions.  In this work we show that any QCA can be put into the form of a PQCA. Our construction reconciles all the non-axiomatic definitions of QCA, showing that they can all simulate one another, and hence that they are all equivalent to the axiomatic definition. This is achieved by defining generalised $n$-dimensional intrinsic simulation, which brings the computer science based concepts of simulation and universality closer to theoretical physics. The result is not only an important simplification of the QCA model, it also plays a key role in the identification of a minimal $n$-dimensional intrinsically universal QCA. 
\end{abstract}

\section{Introduction}\label{sec:introduction}

\subsection{QCA: Importance and Competing Definitions} 

\couic{
The modern axiomatisation of quantum theory in terms of the density matrix formalism was provided by von Neumann in 1955 \cite{NeumannQT}, who also developed the cellular automata (CA) model of computation in 1966 \cite{Neumann},  but he did not bring the two together. Feynman did so in 1986 \cite{FeynmanQC},  just as he was developing the concept of quantum computation (QC). Listed below are the key multidisciplinary motivations for studying QCA, the first two being those of Feynman. 
\VS{\\}\VL{\begin{itemize}}
\VS{-}\VL{\item} \emph{Implementation perspective.} QCA may provide an important path to realistic implementations of QC, mainly because they eliminate the need for an external, classical control over the computation and hence the principal source of decoherence. This is continuously under investigation \cite{BrennenWilliams,LloydQCA,NagajWocjan,Twamley,VollbrechtCirac}.
\VS{-}\VL{\item} \emph{Simulation perspective.} QC was first conceived as a way to efficiently simulate other quantum physical systems. Whilst other applications have been invented since, this still remains a likely and important application of QC. However, it may not be straightforward to encode the theoretical description of a quantum physical system into a QC in a relevant manner, \ie so that the QC can then provide an accurate and efficient simulation. QCA constitute a natural theoretical setting for this purpose, in particular via Quantum Lattice-Gas Automata \cite{Bialynicki-Birula,BoghosianTaylor2,Eakins,LoveBoghosian,MeyerQLGI}.
\VS{-}\VL{\item} \emph{CA perspective.} By their definition (given above), CA are shift-invariant and causal.
CA are therefore a physics-like model of computation (a term coined by Margolus \cite{MargolusPhysics}), as they share some fundamental symmetries of theoretical physics: homogeneity (invariance of physical laws in time and space), causality, and (often) reversibility. Thus it is natural, following Margolus \cite{MargolusQCA}, to study their quantum extensions.
\VS{-}\VL{\item} \emph{Models of computation perspective.} \VL{There are many models of distributed computation (\eg CCS, $\pi$-calculus), but often in such models the idea of space is not directly related to our general understanding, such as our intuitive understanding of relative positions of objects in $3$D space. These models are not adequate for reasoning about simple space-sensitive synchronisation problems, such as `machine self-reproduction' \cite{Codd,Neumann} or the `Firing Squad' problem \cite{MazoyerFiring,MooreFiring}. In contrast, }CA were initially used  to model spatially distributed computation in space \cite{ToffoliMargolusModelling}. Moreover, QCA provide a model of QC, and hence constitute a framework to model and reason about problems in spatially distributed QC.
\VS{-}\VL{\item} \emph{Theoretical physics perspective.} QCA could provide helpful toy models for theoretical physics \cite{LloydQG}. For this purpose it could build bridges between computer science and theoretical physics, as the present paper attempts to for the concept of universality.
\VL{\end{itemize}}
}
Cellular automata (CA), first introduced by von Neumann \cite{Neumann}, consist of an array of identical cells, each of which may take one of a finite number of possible states. The whole array evolves in discrete time steps by iterating a function $G$. This global evolution $G$ is shift-invariant (it acts in the same way everywhere) and local (information cannot be transmitted faster than some fixed number of cells per time step). Because this is a physics-like model of computation \cite{MargolusPhysics}, Feynman \cite{FeynmanQCA}, and later Margolus \cite{MargolusQCA}, suggested 
that quantising this model was important, for two reasons: firstly, because in CA computation occurs without extraneous (unnecessary) control, hence eliminating a source of decoherence; and secondly, because they are a good framework in which to study the quantum simulation of a quantum system. From a computation perspective there are other reasons to study QCA,  such as studying space-sensitive problems in computer science (\eg `machine self-reproduction' \cite{Neumann,MazoyerFiring}) in the quantum setting.  There is also a theoretical physics perspective, where CA are used as toy models of quantum space-time \cite{LloydQG}.

These motivations demonstrate the importance of studying QCA. Once this is acknowledged researchers are faced with an overabundance of competing definitions of QCA.
An examination shows that there are four main approaches to defining QCA: the axiomatic style \cite{ArrighiUCAUSAL,ArrighiLATA,SchumacherWerner}, the multilayer block representation \cite{ArrighiUCAUSAL,PerezCheung}, the two-layer block representation \cite{BrennenWilliams,Karafyllidis,NagajWocjan,Raussendorf,SchumacherWerner,VanDam}, and Partitioned QCA (PQCA) \cite{InokuchiMizoguchi,VanDam,WatrousFOCS}. A natural first questions to consider is whether they are equivalent, and in what sense.

\subsection{QCA: Simulation and Equivalence} \label{subsec:CA}

Probably the most well known CA is Conway's `Game of Life'; a two-dimensional CA which has been shown to be universal for computation in the sense that any Turing Machine (TM) can be encoded within its initial state and then executed by evolution of the CA \cite{winningways}. As TM are generally considered to be a robust definition of `what an algorithm is' in classical computer science, this result could be perceived as providing a conclusion to the topic of CA simulation.  However, this is not the case, as CA do more than just running any algorithm. They run distributed algorithms in a distributed manner, model phenomena together with their spatial structure, and allow the use of the spatial parallelism inherent in the model. These features, modelled by CA and not by TM, are all interesting, and so the concepts of simulation and universality needed be revisited in this context to account for space. This has been done by returning to the original meaning of the word \emph{simulation} \cite{AlbertCulik,Banks,DurandRoka}, namely the ability for one instance of a computational model to simulate other instances of the \emph{same} computational model. The introduction of a partial order on CA via  groupings \cite{MazoyerRapaport}, and subsequent generalisations \cite{OllingerJAC,Theyssier}, have led to elegant and robust definitions of intrinsic simulation. Intrinsic simulation formalises the ability of a CA to simulate another in a space-preserving manner. 
Intuitively this is exactly what is needed to show the equivalence between the various competing definitions of QCA, \ie that they can all simulate each other in a space-preserving manner. The definition of intrinsic simulation has already been translated in the quantum context \cite{ArrighiFI}, however as it stands this is not sufficient to obtain the desired result. In this paper the definition of intrinsic simulation in the quantum context is discussed and developed, before the equivalence between all the various above-mentioned definitions of QCA is tackled. 

\subsection{QCA: Simplification and Universality} \label{subsec:literature}

Intrinsic universality is the ability to intrinsically simulate any other QCA. Here we show that the axiomatic style QCA, the multilayer block representation QCA, the two-layer block representation QCA, and the PQCA are equivalent, entailing that PQCA are intrinsically universal. Here the PQCA is chosen as the prime model as it is
the simplest way to describe a QCA. Therefore, the result developed in this work is also a simplifying one for the field of QCA as a whole. From a theoretical physics perspective, showing that `Partitioned Quantum Cellular Automata are universal' is a statement that `scattering phenomena are universal physical phenomena'.

There are several related results in the CA literature. Several influential works by Morita \etal  emphasise Reversible Partitioned CA universality. For instance, they  provide computation universal Reversible Partitioned CA constructions \cite{MoritaCompUniv1D,MoritaCompUniv2D}, and their ability to simulate any CA in the one-dimensional case is also shown \cite{MoritaIntrinsicUniv1D}. The problem of intrinsically universal Reversible CA (RCA) constructions was tackled by Durand-Lose \cite{Durand-LoseLATIN,Durand-LoseIntrinsic1D}. The difficulty is in having an $n$-dimensional RCA simulate all other $n$-dimensional RCA and not, say, the $(n-1)$-dimensional RCA, otherwise a history-keeping dimension could be used, as in Toffoli \cite{ToffoliConstruction}. Strongly related to this is the work on block representations of RCA by Kari \cite{KariCircuit}.

The QCA-related results are focused on universality. Watrous \cite{WatrousFOCS} proved that QCA are universal in the sense of QTM. Shepherd, Franz and Werner \cite{ShepherdFranz} defined a class of QCA where the scattering unitary $U_i$ changes at each step $i$ (classical control QCA). Universality in the circuit-sense has already been achieved by Van Dam \cite{VanDam}, Cirac and Vollbrecht \cite{VollbrechtCirac}, Nagaj and Wocjan \cite{NagajWocjan}, and Raussendorf \cite{Raussendorf}. In the bounded-size configurations case, circuit universality coincides with intrinsic universality, as noted by Van Dam \cite{VanDam}. Intrinsically universal QCA in the one-dimensional case have also been resolved \cite{ArrighiFI}. 
Finally, a subsequent work, which crucially builds upon the result of this paper, exhibits an $n$-dimensional intrinsically universal QCA \cite{ArrighiNUQCA}. 
Given the crucial role of this in classical CA theory \cite{Durand-LoseEnc}, the issue of intrinsic universality in the $n$-dimensional case needed to be addressed. Having then shown that PQCA, a simple subclass of QCA, are intrinsically universal, it remained to show that there existed a $n$-dimensional PQCA capable of simulating all other $n$-dimensional PQCA for $n>1$, which we show in this paper.

The necessary theoretical background for understanding QCA, and hence the problems addressed in this paper, is provided in section \ref{definitions}.  Intrinsic simulation is discussed and generalised in section \ref{subsecsim}. In section \ref{sec:struc} the various alternative definitions of QCA are shown to be equivalent to the simplest definition, \ie PQCA. Section \ref{sec:discussion} concludes with a discussion and ideas for future directions.

\section{Definitions} \label{definitions}
\subsection{$n$-Dimensional QCA}\label{subsecdef}

This section provides the axiomatic style definitions for $n$-dimensional QCA. 
Configurations hold the basic states of an  entire array of cells, and hence denote the possible basic states of the entire QCA: 
\begin{Def}[Finite configurations]
A \emph{(finite) configuration} $c$ over $\Sigma$ is a function $c: \Z^n \longrightarrow \Sigma$, with 
$(i_1,\ldots,i_n)\longmapsto c(i_1,\ldots,i_n)=c_{i_1\ldots i_n}$, such that there exists a (possibly empty)
finite set $I$ satisfying $(i_1,\ldots,i_n)\notin I\Rightarrow c_{i_1\ldots i_n}=q$, where $q$ is a distinguished \emph{quiescent} state of $\Sigma$.
The set of all finite configurations over $\Sigma$ will be denoted $\mathcal{C}^{\Sigma}_{fin}$.
\end{Def}

Since this work relates to QCA rather than CA, the global state of a QCA can be a superposition of these configurations. 
To construct the separable Hilbert space of superpositions of configurations the set of configurations must be countable. 
Thus finite, unbounded, configurations are considered. The quiescent state of a CA is analogous to the blank symbol of a Turing machine  tape.

\begin{Def}[Superpositions of configurations]\label{superp} 
Let $\mathcal{H}_{\mathcal{C}^{\Sigma}_{fin}}$ be the Hilbert space of configurations. Each finite configuration $c$ is associated with a unit vector $\ket{c}$, such that the family $\pa{\ket{c}}_{c\in\mathcal{C}^{\Sigma}_{fin}}$ is an orthonormal basis of $\mathcal{H}_{\mathcal{C}^{\Sigma}_{fin}}$. A \emph{superposition of configurations} is then a unit vector in $\mathcal{H}_{\mathcal{C}^{\Sigma}_{fin}}$. 
\end{Def}

\begin{Def}[Unitarity]\label{unitarity} A linear operator $G:\mathcal{H}_{\mathcal{C}^{\Sigma}_{fin}}\longrightarrow\mathcal{H}_{\mathcal{C}^{\Sigma}_{fin}}$ is \emph{unitary} if and only if $\{G\ket{c}\,|\,c\in\mathcal{C}^{\Sigma}_{fin}\}$ is an orthonormal basis of $\mathcal{H}_{\mathcal{C}^{\Sigma}_{fin}}.$
\end{Def}

\begin{Def}[Shift-invariance]\label{shift-invariance} 
Consider the shift operation, for $k\in$\\
$\acco{1,\ldots, n}$, which takes configuration $c$ to $c'$ where for all $(i_1,\ldots ,i_n)$ we have $c'_{i_1\ldots i_k \ldots i_n}=c_{i_1\ldots i_k+1 \ldots i_n}$. Let $\sigma_k:\mathcal{H}_{\mathcal{C}^{\Sigma}_{fin}}\longrightarrow\mathcal{H}_{\mathcal{C}^{\Sigma}_{fin}}$ denote its linear extension to a superpositions of configurations. A linear operator $G:\mathcal{H}_{\mathcal{C}^{\Sigma}_{fin}}\longrightarrow\mathcal{H}_{\mathcal{C}^{\Sigma}_{fin}}$ is said to be 
\emph{shift invariant} if and only if $G\sigma_k=\sigma_k G$ for each $k$.
\end{Def}

The following definition captures the  causality of the dynamics. Imposing the condition that the state associated to a cell
(its reduced density matrix) is a function of the neighbouring cells is equivalent to stating that information
propagates at a bounded speed. 
\begin{Def}[Causality]\label{locality} 
A linear operator $G:\mathcal{H}_{\mathcal{C}^{\Sigma}_{fin}}\longrightarrow\mathcal{H}_{\mathcal{C}^{\Sigma}_{fin}}$ is said to be 
 \emph{causal}{} 
if and only if for any
$(i_1,\ldots,i_n)\in\Z_n$,  there exists a function $f$ such that $\rho'|_\joliN = f (\rho|_\joliN)$
for all $\rho$ over $\mathcal{H}_{\mathcal{C}^{\Sigma}_{fin}}$, where:\\
$\joliN=\{i_1,i_1+1\}\times\ldots\times\{i_n,i_n+1\}$, $\rho|_\joliN$ means the restriction of $\rho$ to the neighbourhood $\joliN$ in the sense of the partial trace, and $\rho' = G \rho G^\dagger$.
\end{Def}

In the classical case, the definition is that the letter to be read in some given cell $i$ at time $t+1$ depends only on the state of the cells $i$ to $i+1$ at time $t$. Transposed to a quantum setting, the above definition is obtained. To know the state of cell number $i$, only the states of cells $i$ and $i+1$ before the evolution need be known.

More precisely, this restrictive definition of causality is known in the classical case as a $\frac{1}{2}$-neighbourhood cellular automaton,  because the most natural way to represent such an automaton is to shift the cells by $\frac{1}{2}$ at each step, so that visually the state of a cell depends on the state of the two cells under it. This definition of causality is not restrictive, as by grouping cells into ``supercells'' any CA with an arbitrary finite neighbourhood $\joliN$ can be made into a $\frac{1}{2}$-neighbourhood CA. The same method can be applied to QCA, so this definition of causality holds without loss of generality.
However, the $f$ in the above definition does not directly lead to a constructive definition of a cellular automaton, unlike the local 
transition function in the classical case \cite{ArrighiUCAUSAL}.

\noindent This leads to the definition of an $n$-dimensional QCA.
\begin{Def}\textbf{(QCA)}\label{def:qca} 
An $n$-dimensional quantum cellular automaton (QCA) is an operator $G:\mathcal{H}_{\mathcal{C}^{\Sigma}_f}\longrightarrow\mathcal{H}_{\mathcal{C}^{\Sigma}_f}$
which is unitary, shift-invariant and causal.
\end{Def}
Whilst this is clearly the natural, axiomatic definition QCA, clearly stems from an equivalent definition in the literature, phrased in terms of homomorphism of a $C^*$-algebra \cite{SchumacherWerner}. However, it remains a non-constructive definition and in this sense it can be compared to the Curtis-Hedlund \cite{Hedlund} definition of CA as the set of continuous, shift-invariant functions. These definitions characterise (Q)CA via the global, composable properties that they must have; but they do not provide an operational, hands-on description of their dynamics. clearly stems from an equivalent definition in the literature, phrased in terms of homomorphism of a $C^*$-algebra \cite{SchumacherWerner}. This work aims at further simplifying those mathematical objects, down to PQCA.

\subsection{Multilayer Block Representation}

The axiomatic style definition of QCA remains somewhat abstract and mathematical.
A central tool and concept in this paper is that of a (multilayer) block representation of QCA. Intuitively, we say that a QCA $G$ admits a block representation when it can be expressed as blocks, \ie local unitaries, composed in space (via the tensor product) and time (via operator composition), thereby forming a finite-depth quantum circuit infinitely repeating across space. The structure theorem given in previous work \cite{ArrighiUCAUSAL} states that any QCA can in fact be represented in such a way:

\begin{Th}[$n$-dimensional QCA multilayer block representation]\label{th:multilayers}~\\
Let $G$ be an $n$-dimensional QCA with alphabet $\Sigma$. Let $E$ be an isometry from $\joliH_\Sigma\to\joliH_\Sigma\otimes\joliH_\Sigma$ such that $E\ket{\psi_x}=\ket{q}\otimes\ket{\psi_x}$. This mapping can be trivially extended to whole configurations, yielding a mapping $E:\joliH_{C^{\Sigma}_f}\to\joliH_{C^{\Sigma^2}_f}$. There then exists an $n$-dimensional QCA $H$ on alphabet $\Sigma^2$, such that $HE=EG$, and $H$ admits an $2^n$-layer block representation. Moreover $H$ is of the form 
\begin{align}
H=(\bigotimes S)(\prod K_x) \label{eq:luqca}
\end{align}
where:
\begin{itemize}
\item $(K_x)$ is a collection of commuting unitary operators all identical up to shift, each localised upon each neighbourhood $\joliN_x$;
\item $S$ is the swap gate over $\joliH_\Sigma\otimes\joliH_\Sigma$, hence localised upon each node $x$.
\end{itemize}
\end{Th}

This theorem therefore bridges the gap between the axiomatic style definition of QCA and the operational descriptions of QCA. Again, it should be compared with the Curtis-Hedlund \cite{Hedlund} theorem, which shows the equivalence between the axiomatic definition of CA and the more operational, standard definition, with a local function applied synchronously across space. One can argue that the form given in Eqn.~\ref{eq:luqca} is not that simple. A contribution of this paper is to simplify it down to PQCA.

Amongst the operational definitions of QCA listed in section \ref{sec:introduction},
only that of Perez-Delgado and Cheung \cite{PerezCheung} is not two-layer. 
They directly state, after some interesting informal arguments, that QCA are of a form similar
to that given in Eqn.~\ref{eq:luqca}.

In other words, this theorem says that starting from an axiomatic definition of QCA, such as Shumacher and Werner's \cite{SchumacherWerner}, one can derive a
circuit-like structure for $n$-dimensional QCA, thereby extending their result to $n$ dimensions. It also
shows that operational definitions \cite{PerezCheung} can be given a rigorous axiomatics. 
It follows that the definitions of P\'erez-Delgado and Cheung \cite{PerezCheung} and Shumacher and Werner \cite{SchumacherWerner} are actually equivalent, up to ancillary cells.

Therefore the axiomatic definition of QCA given in section \ref{subsecdef} is equivalent to a multilayer block representation.
There are, however, several other definitions of QCA, \ie two-layer block representations and PQCA. 
The aim is to now show that all definitions of QCA  can be reconciled via
intrinsic simulation. A quantum version of intrinsic simulation has already been developed \cite{ArrighiFI},
but only for one-dimensional QCA, and it is not general enough to state the required equivalence. 
This difficulty is addressed in the next section, where a new concept of intrinsic simulation for $n$-dimensional QCA
is developed with the required properties.

\section{Intrinsic Simulation of $n$-Dimensional QCA} \label{subsecsim}

Intrinsic simulation of one CA by another was discussed informally in section \ref{subsec:CA}. A
pedagogical discussion in the classical case was given by Ollinger \cite{OllingerJAC}, and
quantised intrinsic simulation has been formalised in the one-dimensional case \cite{ArrighiFI}. 
This definition is extended to $n$-dimensions (and relaxed, see details below) here.
The potential use of this concept in theoretical physics is also discussed.

Intuitively, `$G$ simulates $H$' is shown by translating the contents of each cell of $H$ into cells of $G$,
running $G$, and then reversing the translation;  this three step process amounts to running $H$. 
This translation should be simple (it should not provide a ``hidden'' way to compute $G$),  should preserve the topology (each cell of $H$ is encoded into cells of $G$ in a way which preserves neighbours), and  should be faithful (no information should be lost in translation). This latter requirement relates to the \emph{isometry} property of quantum theory, \ie an inner product preserving evolution with $Enc^\dagger Enc=\mathbb{I}$. This same requirement agrees with the translation being a physical process. The following definitions are thus derived.

\begin{Def}[Isometric coding]\label{isomcode} 
Consider $\Sigma_G$ and $\Sigma_H$, two alphabets with distinguished quiescent states $q_G$ and $q_H$, and such that $|\Sigma_H|\leq|\Sigma_G|$. Consider $\mathcal{H}_{\Sigma_G}$ and $\mathcal{H}_{\Sigma_H}$ the Hilbert spaces having these alphabets as their basis, and $\mathcal{H}_{\mathcal{C}_{fin}^{G}}$, $\mathcal{H}_{\mathcal{C}_{fin}^{H}}$ the Hilbert spaces of finite configurations over these alphabets.\\
Let $E$ be an isometric linear map from $\mathcal{H}_{\Sigma_H}$ to $\mathcal{H}_{\Sigma_G}$ which preserves quiescence, \ie such that $E\ket{q_H}=\ket{q_G}$. It trivially extends into an isometric linear map $Enc=(\bigotimes_{\mathbb{Z}^n} E)$ from $\mathcal{H}_{\mathcal{C}_{fin}^{H}}$ into $\mathcal{H}_{\mathcal{C}_{fin}^{G}}$, which we call an isometric encoding.\\
Let $D$ be an isometric linear map from $\mathcal{H}_{\Sigma_G}$ to $\mathcal{H}_{\Sigma_H}\otimes\mathcal{H}_{\Sigma_G}$ which also preserves quiescence, in the sense that $D\ket{q_G}=\ket{q_H}\otimes\ket{q_G}$. It trivially extends into an isometric linear map $Dec=(\bigotimes_{\mathbb{Z}^n} D)$ from $\mathcal{H}_{\mathcal{C}_{fin}^{G}}$ into $\mathcal{H}_{\mathcal{C}_{fin}^{H}}\otimes\mathcal{H}_{\mathcal{C}_{fin}^{G}}$, which we call an isometric decoding.\\
The isometries $E$ and $D$ define an isometric coding if the following condition is satisfied:\\
$\forall \ket{\psi}\in \mathcal{H}_{\mathcal{C}_{fin}^{H}},\,\exists \ket{\phi}\in \mathcal{H}_{\mathcal{C}_{fin}^{G}}\quad/\quad\ket{\psi}\otimes\ket{\phi}=Dec\pa{Enc \ket{\psi}}.$
\end{Def}

(Here $Dec$ is understood to morally be an inverse function of $Enc$, but some garbage $\ket{\phi}$ may be omitted.)

\begin{Def}[Direct simulation]\label{directsim}
Consider $\Sigma_G$ and $\Sigma_H$, two alphabets with distinguished quiescent states $q_G$ and $q_H$, and two QCA $G$ and $H$ over these alphabets. We say that $G$ directly simulates $H$, if and only if there exists an isometric coding such that\\
$\forall i\in\mathbb{N},\,\forall \ket{\psi}\in \mathcal{H}_{\mathcal{C}_{fin}^{H}},\,\exists \ket{\phi}\in \mathcal{H}_{\mathcal{C}_{fin}^{G}}\quad/\quad (G^i\ket{\psi})\otimes\ket{\phi}=Dec \pa{{H}^i\pa{Enc \ket{\psi}}}.$
\end{Def}

\noindent Unfortunately this is not enough for intrinsic simulation, as it implies that $|\Sigma_H| = |\Sigma_G|$. It is often desirable  that $G$ simulates $H$ even though the translation:\\
- takes several cells of $H$ into several cells of $G$;\\
- demands several steps of $G$ in order to simulate several steps of $H$.\\
Hence the grouping of cells is required.
\begin{Def}[Grouping]\label{def:packmap} 
Let $G$ be an $n$-dimensional QCA over alphabet $\Sigma$. Let $s$ and $t$ be two integers, $q'$ a word in $\Sigma'=\Sigma^{s^n}$. Consider the iterate global evolution $G^t$ up to a grouping of each hypercube of $s^n$ adjacent cells into one supercell. If this operator can be considered to be a QCA $G'$ over $\Sigma'$ with quiescent symbol $q'$, then we say that $G'$ is an $(s,t,q')$-grouping of $G$.
\end{Def}

A natural way to continue would be to define an intrinsically universal QCA. However, due to the continuity of $\mathcal{H}$, this approximation
can only be up to $\epsilon$. In the companion paper to this we provide a intrinsically  universal instance of a QCA with a bound on the finite error \cite{ArrighiSimple}.

\begin{Def}[Intrinsic simulation]\label{def:intsim}
Consider $\Sigma_G$ and $\Sigma_H$, two alphabets with distinguished quiescent states $q_G$ and $q_H$, and two QCA $G$ and $H$ over these alphabets. We say that $G$ intrinsically simulates $H$ if and only if there exists $G'$, some grouping of $G$, and $H'$, some grouping of $H$, such that $G'$ directly simulates $H'$.
\end{Def}

In other words, $G$ intrinsically simulates $H$ if and only if there exists some isometry $E$ which translates supercells of $H$ into supercells of $G$, such that if $G$ is iterated and then translated back, the whole process is equivalent to an iteration of $H$. {This understanding is shown schematically in Fig.~\ref{IntrinsicSim}.
\begin{figure}
\centering
\includegraphics[scale=1, clip=true, trim=0cm 0cm 0cm 0cm]{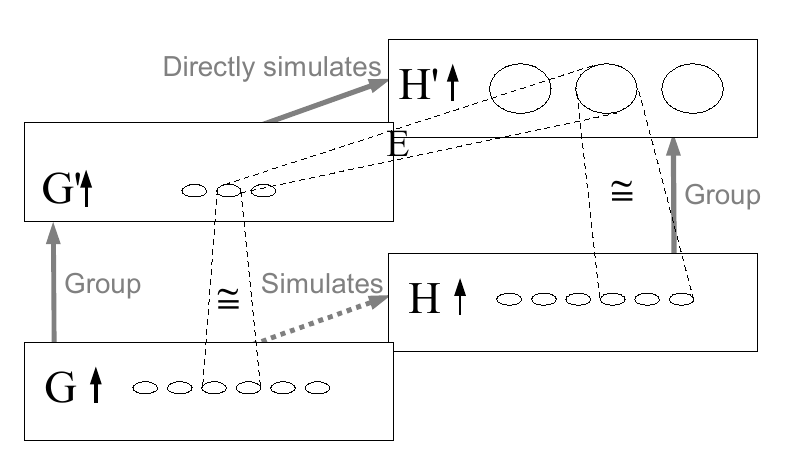}
\caption{The concept of intrinsic simulation made formal.\label{IntrinsicSim}}
\end{figure}
} 

Compared with previous work \cite{ArrighiFI}, the concept of intrinsic simulation has been modified to allow the grouping
in Fig.~\ref{IntrinsicSim} on the simulated QCA side, and this variation is important to Thm.~\ref{th:pqca}.
This is analogous to the classical case \cite{Theyssier}. 

A natural way to follow would be to define the notion of an intrinsically universal QCA. However due to the continuous nature of the underlying Hilbert spaces, no QCA can be intrinsically universal in an exact sense. We can only hope to have a `dense' QCA, \ie one which can simulate any other up to some precision $\epsilon$, which can then be made arbitrarily small. In \cite{ArrighiNUQCA}, such an $n$-dimensional intrinsically universal QCA construction is indeed provided, together with bounds on $\epsilon$. Notice that the latter result \cite{ArrighiNUQCA} crucially relies upon the fact that PQCA are intrinsically universal.
 
\VL{
\section{Constructions} \label{sec:struc}
Now that an appropriate notion of intrinsic simulation has been developed, the problem of showing an equivalence between the
different operational definitions of QCA is addressed here.

\subsection{Down to Two Layers: Block QCA}\label{subsec:twolayers}
Quantisations of block representations of CA are generally presented as two-layer; \cf \cite{BrennenWilliams,Karafyllidis,NagajWocjan,Raussendorf,SchumacherWerner,VanDam}. This is captured by the definition of a Block QCA (BQCA), where $\mathcal{H}^{\otimes 2^n}$ is $\mathcal{H}\otimes\ldots\otimes\mathcal{H}$, repeated $2^n$ times:
\begin{Def}[BQCA]\label{def:bqca}
A block $n$-dimensional quantum cellular automaton (BQCA) is defined by two unitary operators $U_0$ and $U_1$ such that $U_i:\mathcal{H}_{\Sigma}^{\otimes 2^n}\longrightarrow\mathcal{H}_{\Sigma}^{\otimes 2^n}$, and $U_i\ket{qq\ldots qq}=\ket{qq\ldots qq}$, \ie each takes $2^n$ cells into $2^n$ cells and preserves quiescence. Consider $G_i=(\bigotimes_{2\mathbb{Z}^n} U_i)$ the operator over $\mathcal{H}$. The induced global evolution is $G_0$ at odd time steps, and $\sigma G_1$ at even time steps, where $\sigma$ is a translation by one in all directions (Fig.~\ref{fig:structureBQCA}). 
\end{Def}

\begin{figure}
\centering
\includegraphics[scale=1.1, clip=true, trim=0cm 0cm 0cm 0cm]{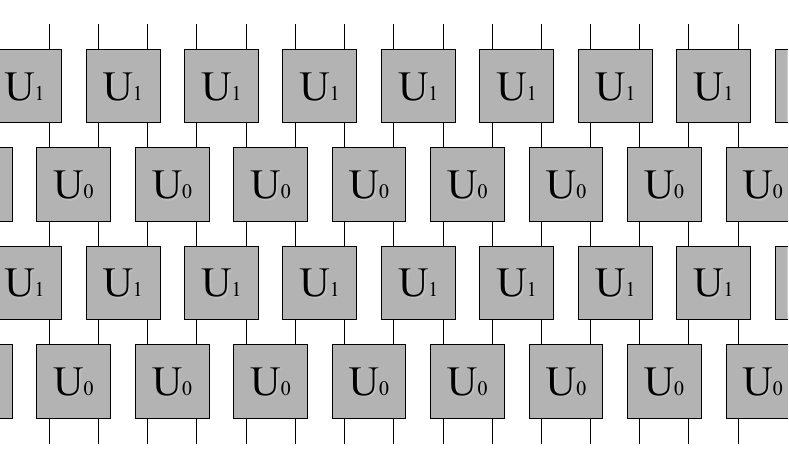}
\caption{BQCA.\label{fig:structureBQCA} The elementary unitary evolutions $U_0$ and $U_1$ are alternated repeatedly as shown, in 1D.}
\end{figure}

Showing the equivalence of the QCA and BQCA axiomatics is not trivial. 
In one direction this is simple, as 
BQCA are unitary, causal, and shift-invariant, and hence fall under the axiomatics 
and Thm.~\ref{th:multilayers}  (strictly speaking we need to group each hypercube of $2^n$ adjacent cells into a supercell,
see Def.~\ref{def:packmap}.) 
However, there are several factors to consider regarding the ability of BQCA to simulate any QCA, which are now addressed. 

In the form given by Thm.~\ref{th:multilayers}, each cell $x$ at time $t$ is successively involved in $2^n$ computations governed by a local unitary $K$, whose aim is to compute the next state of a cell within a radius $\frac{1}{2}$ from $x$ at time $t+1$. In two dimensions, a cell $x$ uses the cells West, North-West and North to work out its North-West successor, and then the cells North, North-East, East of it to compute the North-East successor (Similarly for the South-East and the South-West successors). To mimic this with a BQCA,  each original cell can be encoded into four cells, arranged so that the original cell $x$ starts in the North-West quadrant of the four cells. The first layer of the BQCA  applies the local unitary $K$ to compute the North-West successor of $x$. The second layer of the BQCA moves the original cell $x$ in the North-West quadrant. Each full application of the evolution of the BQCA corresponds only to one layer $(\bigotimes K)$, hence it will take four steps for this BQCA to simulate one step of the QCA. Fig.~\ref{fig:2layersSketch} shows a sketch of the method used. 
\begin{figure}[h!]
\centering
\includegraphics[scale=.75, clip=true, trim=0cm 0cm 0cm 0cm]{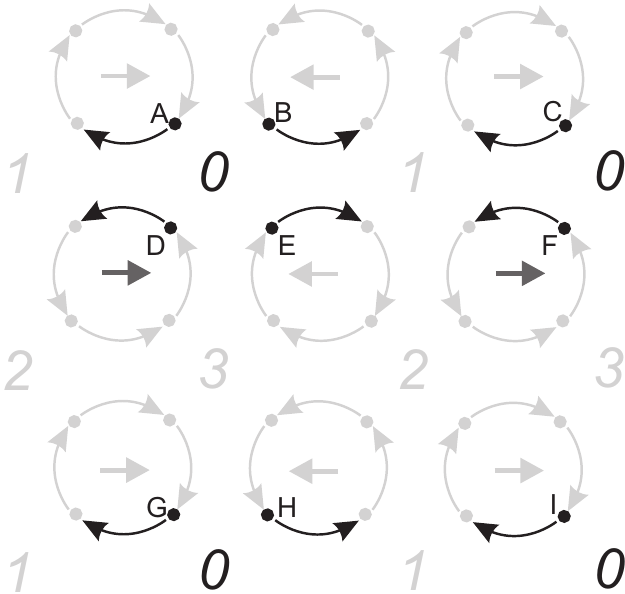}
\caption{Sketch of a BQCA simulating a QCA.\label{fig:2layersSketch} The original cell $x$ is coded into four cells, at the centre ($E$). It starts by considering the North-West as at time $0$ it will compute its North-West successor, and then move clockwise. At time $1$ it will compute its North-East successor, etc.}
\end{figure}

There are some considerations to be discussed. When cell $x$ is turning clockwise in the example, the cell to its North is turning anticlockwise. Hence we need some ancillary data coding for the path to be taken by the original cell $x$ within the four coding cells. 
Also, Thm.~\ref{th:multilayers} finishes with a $Swap$ between the `computed tape', where the results have been stored, and the `uncomputed tape', (\ie what remains of the original cell after having computed all of its successors) which is not shown in the sketch.
Hence the number of layers of $K$ computed so far has to be tracked, so that the $Swap$ occurs at the appropriate step. The $Swap$ also needs to know where the results have been stored in order to move them correctly. All of this has to be arranged spatially and efficiently, and one such method is shown in Figs. \ref{fig:2layersBig} and \ref{fig:2layersOps}. 
\begin{figure}[h!]
\centering
\VS{\includegraphics[scale=.5, clip=true, trim=0cm 0cm 0cm 0cm]{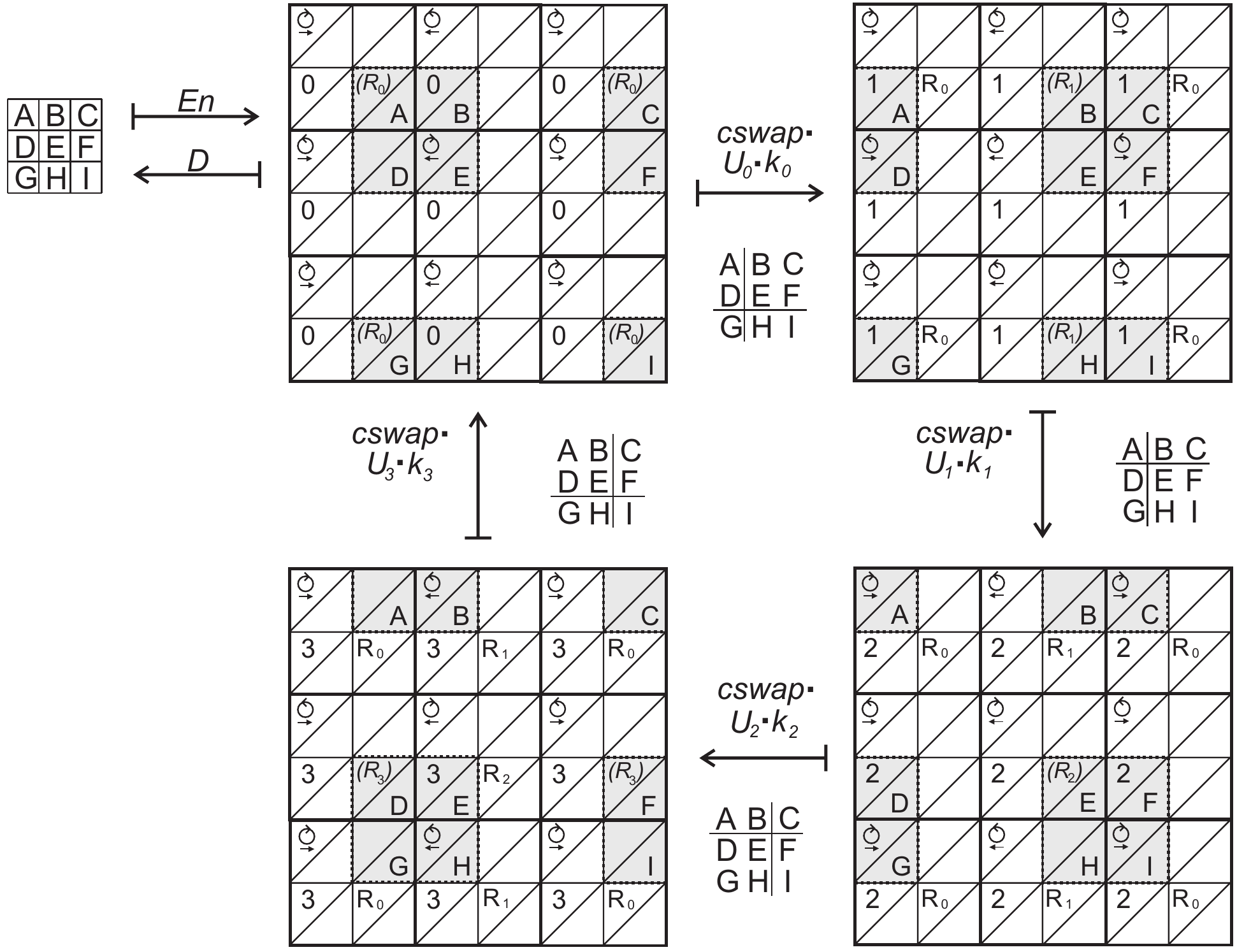}}
{\includegraphics[scale=.6, clip=true, trim=0cm 0cm 0cm 0cm]{img/QCA2.pdf}}
\caption{BQCA simulating a QCA. The grey areas denote the neighbourhood where the action of $k_x$, the first layer of the BQCA, will be significant -- \ie a group of four cells where it will perform a $K_x$ operation to work out a successor. Where this successor will be stored is indicated by $(R_x)$. At the next step $R_x$ has appeared, and the registers have been reshuffled due to the second layer of the BQCA, which acts according to the rotation-direction mark. The second layer also increases the clock count and includes the final swapping step, which only happens at time $3$. There it  ensures that $R_0$ becomes $A$, $R_1$ becomes $B$, etc. Which registers are to be swapped with one another can be calculated from the rotation and arrow marks. Each step is made formal by Fig.~\ref{fig:2layersOps}.\label{fig:2layersBig}}
\end{figure}
\begin{figure}[h!]
\centering
\includegraphics[scale=.7, clip=true, trim=0cm 0cm 0cm 0cm]{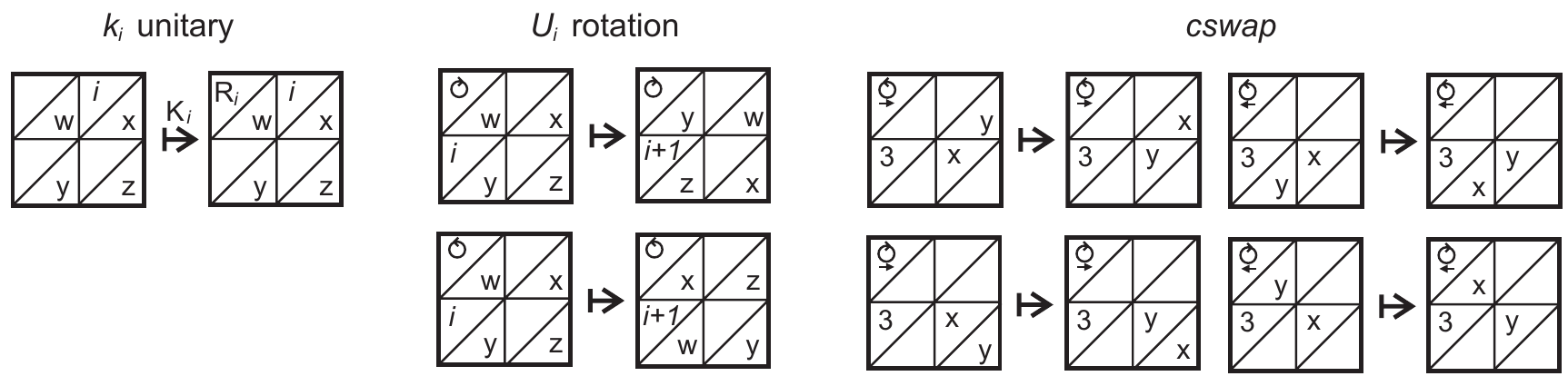}
\caption{Operations used in Fig.~\ref{fig:2layersBig}.  $k$ applies a $K$ operation whenever some data is present (data carries an extra bit to distinguish it from $\ket{q}$, say). The $U$ operation reshuffles the data by rotating it in the direction given by the indicator in the top left (clockwise or anticlockwise), and increments the index counter. Finally, $cswap$ acts as the identity in all cases except when the index is 3, when it swaps the result of the computations with the data, ready for the next round.\label{fig:2layersOps}}
\end{figure}

BQCA can therefore simulate QCA up to a relatively simple encoding, using blocks of
four cells. This explains the need for grouping on the simulated QCA side in the revised quantised intrinsic simulation, as in Fig.~\ref{IntrinsicSim}.
Encoding groups of cells rather than individual cells is also required for the PQCA discussion (\videinfra). 
This encoding is given for two dimensions, but the construct clearly generalises to $n$-dimensions. Hence QCA (Def.~\ref{def:qca})
provide a rigorous axiomatics for BQCA (Def.~\ref{def:bqca}), and BQCA provide a convenient operational description of QCA. We have shown that:
\begin{Cl}[BQCA are universal]\label{th:bqca}
Given any $n$-dimensional QCA $H$, there exists an $n$-dimensional BQCA $G$ which simulates $H$.
\end{Cl}

\subsection{Down to One Scattering Unitary: PQCA}\label{subsec:onescattering}

\begin{figure}[h]
\centering
\includegraphics[scale=.9, clip=true, trim=0cm 0cm 0cm 0cm]{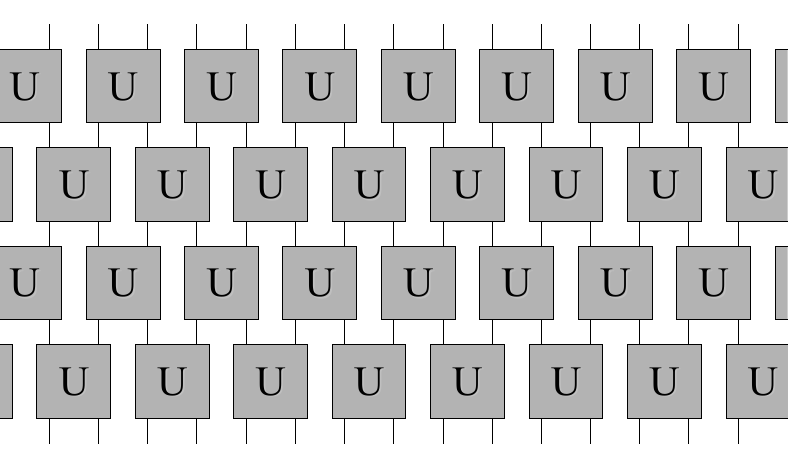}
\caption{\label{fig:structure} Partitioned one-dimensional QCA with scattering unitary $U$. Each line represents a quantum system, in this case a whole cell. Each square represents a scattering unitary $U$ which is applied to two cells. Time flows upwards.}
\end{figure}

Quantisations of partitioned representations of CA are given in several works \cite{InokuchiMizoguchi,VanDam,WatrousFOCS}. 
These constitute the simplest approach to defining QCA.
It is therefore interesting to consider whether QCA (as in Def.~\ref{def:qca}) provide a rigorous axiomatics for PQCA, and if PQCA provide a convenient operational description of QCA. A PQCA is essentially a BQCA where the two layers apply the same unitary operation, shifted appropriately.
\begin{Def}[PQCA]\label{def:pqca}
A partitioned  $n$-dimensional quantum cellular automaton (PQCA) is defined by a scattering unitary operator $U$ such that $U:\mathcal{H}_{\Sigma}^{\otimes 2^n}\longrightarrow\mathcal{H}_{\Sigma}^{\otimes 2^n}$, and $U\ket{qq\ldots qq}=\ket{qq\ldots qq}$, \ie that takes
a hypercube of $2^n$ cells into a hypercube of $2^n$ cells and preserve quiescence. Consider $G=(\bigotimes_{2\mathbb{Z}^n} U)$, the operator over $\mathcal{H}$. The induced global evolution is $G$ at odd time steps, and $\sigma G$ at even time steps, where $\sigma$ is a translation by one in all directions (Fig.~\ref{fig:structure}).
\end{Def}

{
\begin{figure}[h!]
\centering
\includegraphics[scale=.71, clip=true, trim=0cm 0cm 0cm 0cm]{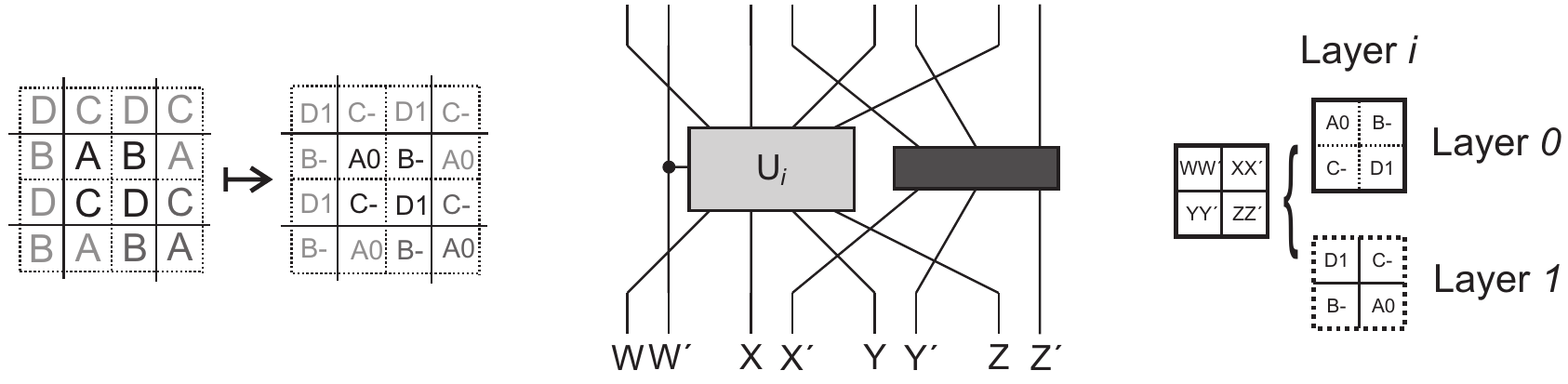}
\caption{PQCA simulating a BQCA. The QCA is decorated with control qubits following a simple encoding procedure (\emph{left}), which allow the scattering unitary $U$ (\emph{centre}) to act as either $U_0$ or $U_1$, according to the layer (\emph{right}). The black box can be any unitary. \label{fig:PQCAbig}}
\end{figure}
}\VS{
\begin{figure}[h!]
\centering
\includegraphics[scale=.7, clip=true, trim=0cm 0cm 0cm 0cm]{img/QCA1new.pdf}
\caption{PQCA simulating a BQCA. The QCA is decorated with control qubits following a simple encoding procedure (\emph{left}), which allow the scattering unitary $U$ (\emph{centre}) to act as either $U_0$ or $U_1$, according to the layer (\emph{right}). The black box can be any unitary. \label{fig:PQCAbig}}
\end{figure}
}

Following previous results  (section \ref{subsec:twolayers}), it is only necessary to show that PQCA can simulate BQCA. Both PQCA and BQCA are two-layer; the only difference is that for BQCA those two layers may be different (\eg compare Figs.~\ref{fig:structure} and \ref{fig:structureBQCA}),
whereas for PQCA there is only a single scattering unitary. So a $U$-defined PQCA, with a $U$ capable of performing $U_0$ and $U_1$ alternatively as controlled by some ancillary suffices. This has been shown for one dimension \cite{ArrighiFI} and is given here for two dimensions in Fig.~\ref{fig:PQCAbig}. It is clear that the construct given here generalises to $n$-dimensions.\\

\begin{Cl}[PQCA are universal]\label{th:pqca}
Given any $n$-dimensional QCA $H$, there exists an $n$-dimensional PQCA $G$ which simulates $H$.
\end{Cl}

It can therefore be concluded that PQCA are the most canonical and general operational description of QCA. More generally, by 
showing here that the various definitions of QCA available  \cite{Watrous,VanDam,BrennenWilliams,NagajWocjan,Raussendorf,Karafyllidis,InokuchiMizoguchi,PerezCheung} are equivalent, we argue that a well-axiomatised, concrete, and operational $n$-dimensional QCA is now available.
}

\section{Conclusion}\label{sec:discussion}
The main contribution of this paper is to show that PQCA are intrinsically universal. There are several consequences, summarised here:
\begin{itemize}
\item The construction shows that all the non-axiomatic definitions of QCA \cite{BrennenWilliams,InokuchiMizoguchi,Karafyllidis,NagajWocjan,PerezCheung,Raussendorf,VanDam,WatrousFOCS} are equivalent to one another and to the axiomatic definition, \ie they all simulate each other. Therefore the concept of $n$-dimensional QCA is well-axiomatised, concrete, and operational. 
\item The QCA model is simplified, \ie without loss of generality a QCA can be assumed to be a PQCA (see Def.~\ref{def:pqca}).
\item The identification of an $n>1$-dimensional intrinsically universal QCA is greatly simplified, as it suffices to isolate one $n>1$-dimensional PQCA capable of simulating any other $n$-dimensional PQCA. This is achieved in a subsequent work \cite{ArrighiNUQCA}.
\end{itemize}
Another contribution of this paper is to define and promote $n$-dimensional intrinsic universality as a useful concept, {\em per se}.  This has to be evaluated within a wider context, explained below.\\

\noindent {\em Universality: From Theoretical Computer Science to Theoretical Physics.}\\
\indent The study of QC has so far aimed to address the issues related to the physical nature of computing, and 
over the last twenty years there have been a number of quantisations of the classical models of computation, and novel results on the complexity of the tasks that can be encoded in these models. It could be said that theoretical physics has aided theoretical computer science via this path. It is, however, not unlikely that the reverse path could also be productive. This would be part of a bigger trend where theoretical physics departs from looking at `matter' (particles interacting, scattering, forces, etc.) and seeks to look at `information' (entropy, observation, information exchanges between systems, etc.), in an attempt to clarify its own concepts. An example of this is the huge impact that quantum information theory has had on the understanding of foundational concepts such as entanglement \cite{DurVidalCirac} and decoherence \cite{PazZurek}. A computer science based approach could help to understand basic physical principles, not only in terms of `information', but also in terms of the `dynamics of information', \ie information processing. 


Universality, among the many  concepts in computer science, is a simplifying methodology in this respect. For example, if the problem being studied crucially involves some idea of interaction, universality makes it possible to cast it in terms of information exchanges \emph{together} with some universal information processing. This paper presents an attempt to export universality as a tool for application in theoretical physics; and hence is a small step towards the goal of finding and understanding what is a \emph{universal physical phenomenon}, within some simplified mechanics. Let us refine this idea:
\begin{itemize}
\item Firstly, we want to be able to simulate each physical object independently in its own space. Hence a universal physical phenomenon should be some elementary unit of computation that can be combined to form a 3D network, accounting for space and interactions across space satisfactorily. (The classical universal TM, on the other hand, does not simulate objects in their own space.) 
\item Secondly, we want to be able to simulate simple physical objects in an efficient manner, even if they are quantum. Hence a universal physical phenomenon should be a universal model of \emph{quantum} computation. (The classical universal TM is slow at simulating simple quantum phenomena, which suggests that it is not rich enough).
\end{itemize}
The work that has been presented in this paper formalises an idea of universality which fits both these criteria, namely intrinsic universality over QCA.\\

\noindent {\em A Physical Interpretation.} \\
\indent With this understanding, the intrinsic universality of PQCA developed here could be given a physical interpretation. QCA, as seen through their axiomatic definition (Def.~\ref{def:qca}), are synonymous with discrete-time, discrete-space quantum mechanics (together with some extra assumptions such as translation-invariance and finite-density of information). Stating that discrete-time, discrete-space quantum mechanical evolutions can, without loss of generality, be assumed to be of the form illustrated in Fig.~\ref{fig:structure}, amounts to the statement that `scattering phenomena are universal physical phenomena'. In this sense, the result leads to an understanding of the links between the axiomatic, top-down principles approach to theoretical physics, and the more bottom-up study of the scattering of particles.

\VS{\vspace{-4mm}}\section*{Acknowledgements}
\VS{\vspace{-2mm}}The authors would like to thank J\'er\^ome Durand-Lose, Jacques Mazoyer, Nicolas Ollinger, Guillaume Theyssier and Philippe Jorrand.

\bibliography{../../../Bibliography/biblio}
\bibliographystyle{plain}

\end{document}

%% file: braket.tex
\def\ket#1{\mathinner{|{#1}\rangle}}

{\catcode`\|=\active 
  \gdef\Braket#1{\begingroup \mathcode`\|32768\let|\BraVert\left<{#1}\right>\endgroup}
}
\def\BraVert{\egroup\,\mid@vertical\,\bgroup}
{\catcode`\|=\active
  \gdef\set#1{\mathinner{\lbrace\,{\mathcode`\|"8000\let|\midvert #1}\,\rbrace}}
  \gdef\Set#1{\left\{\:{\mathcode`\|"8000\let|\SetVert #1}\:\right\}}}
\def\midvert{\egroup\mid\bgroup}
\def\SetVert{\egroup\;\mid@vertical\;\bgroup}

%% file: pqca.bbl
\begin{thebibliography}{10}

\bibitem{AlbertCulik}
J.~Albert and K.~Culik.
\newblock {A simple universal cellular automaton and its one-way and totalistic
  version}.
\newblock {\em Complex Systems}, 1:1--16, 1987.

\bibitem{ArrighiFI}
P.~Arrighi, R.~Fargetton, and Z.~Wang.
\newblock {Intrinsically universal one-dimensional quantum cellular automata in
  two flavours}.
\newblock {\em Fundamenta Informaticae}, 21:1001--1035, 2009.

\bibitem{ArrighiNUQCA}
P.~Arrighi and J.~Grattage.
\newblock {Intrinsically universal $n$-dimensional quantum cellular automata}.
\newblock Journal version of refereed conference proceedings, submitted. ArXiv
  preprint: arXiv:0907.3827, 2009.

\bibitem{ArrighiSimple}
P.~Arrighi and J.~Grattage.
\newblock {A Simple $n$-Dimensional Intrinsically Universal Quantum Cellular
  Automaton}.
\newblock {\em Language and Automata Theory and Applications, Lecture Notes in
  Computer Science}, 6031:70--81, 2010.

\bibitem{ArrighiUCAUSAL}
P.~Arrighi, V.~Nesme, and R.~Werner.
\newblock {Unitarity plus causality implies localizability}.
\newblock {\em {QIP 2010 and Journal of Computer and System Sciences, ArXiv
  preprint: arXiv:0711.3975}}, 2010.

\bibitem{ArrighiLATA}
P.~Arrighi, V.~Nesme, and R.~F. Werner.
\newblock {Quantum cellular automata over finite, unbounded configurations}.
\newblock In {\em Proceedings of MFCS, Lecture Notes in Computer Science},
  volume 5196, pages 64--75. Springer, 2008.

\bibitem{Banks}
E.~R. Banks.
\newblock {Universality in cellular automata}.
\newblock In {\em SWAT '70: Proceedings of the 11th Annual Symposium on
  Switching and Automata Theory (SWAT 1970)}, pages 194--215, Washington, DC,
  USA, 1970. IEEE Computer Society.

\bibitem{winningways}
E.R. Berlekamp, J.H. Conway, and R.K. Guy.
\newblock {\em {Winning ways for your mathematical plays}}.
\newblock AK Peters, Ltd., 2003.

\bibitem{BrennenWilliams}
G.~K. Brennen and J.~E. Williams.
\newblock {Entanglement dynamics in one-dimensional quantum cellular automata}.
\newblock {\em Phys. Rev. A}, 68(4):042311, Oct 2003.

\bibitem{DurVidalCirac}
W.~D{\"u}r, G.~Vidal, and J.~I. Cirac.
\newblock {Three qubits can be entangled in two inequivalent ways}.
\newblock {\em Phys. Rev. A}, 62:062314, 2000.

\bibitem{DurandRoka}
B.~Durand and Z.~Roka.
\newblock {The Game of Life: universality revisited Research Report 98-01}.
\newblock Technical report, Ecole Normale Suprieure de Lyon, 1998.

\bibitem{Durand-LoseLATIN}
J.~O. Durand-Lose.
\newblock {Reversible cellular automaton able to simulate any other reversible
  one using partitioning automata}.
\newblock In {\em In {LATIN'95}: Theoretical Informatics, number 911 in Lecture
  Notes in Computer Science}, pages 230--244. Springer, 1995.

\bibitem{Durand-LoseIntrinsic1D}
J.~O. Durand-Lose.
\newblock {Intrinsic universality of a 1-dimensional reversible cellular
  automaton}.
\newblock In {\em Proceedings of STACS 97, Lecture Notes in Computer Science},
  page 439. Springer, 1997.

\bibitem{Durand-LoseEnc}
J.~O. Durand-Lose.
\newblock {Universality of Cellular Automata}.
\newblock In {\em Encyclopedia of Complexity and System Science}, page~22.
  Springer, 2008.

\bibitem{FeynmanQCA}
R.~P. Feynman.
\newblock {Quantum mechanical computers}.
\newblock {\em Foundations of Physics (Historical Archive)}, 16(6):507--531,
  1986.

\bibitem{Hedlund}
G.~A. Hedlund.
\newblock {Endomorphisms and automorphisms of the shift dynamical system}.
\newblock {\em Math. Systems Theory}, 3:320--375, 1969.

\bibitem{InokuchiMizoguchi}
S.~Inokuchi and Y.~Mizoguchi.
\newblock {Generalized partitioned quantum cellular automata and quantization
  of classical CA}.
\newblock {\em International Journal of Unconventional Computing, ArXiv
  preprint: quant-ph/0312102}, 1:149--160, 2005.

\bibitem{Karafyllidis}
I.~G. Karafyllidis.
\newblock {Definition and evolution of quantum cellular automata with two
  qubits per cell}.
\newblock {\em Journal reference: Phys. Rev. A}, 70:044301, 2004.

\bibitem{KariCircuit}
J.~Kari.
\newblock {On the circuit depth of structurally reversible cellular automata}.
\newblock {\em Fundamenta Informaticae}, 38(1-2):93--107, 1999.

\bibitem{LloydQG}
S.~Lloyd.
\newblock {A theory of quantum gravity based on quantum computation}.
\newblock ArXiv preprint: quant-ph/0501135, 2005.

\bibitem{MargolusPhysics}
N.~Margolus.
\newblock {Physics-like models of computation}.
\newblock {\em Physica D: Nonlinear Phenomena}, 10(1-2), 1984.

\bibitem{MargolusQCA}
N.~Margolus.
\newblock {Parallel quantum computation}.
\newblock In {\em Complexity, Entropy, and the Physics of Information: The
  Proceedings of the 1988 Workshop on Complexity, Entropy, and the Physics of
  Information Held May-June, 1989, in Santa Fe, New Mexico}, page 273. Perseus
  Books, 1990.

\bibitem{MazoyerFiring}
J.~Mazoyer.
\newblock {A Six-State Minimal Time Solution to the Firing Squad
  Synchronization Problem}.
\newblock {\em Theoretical Computer Science}, 50:183--238, 1987.

\bibitem{MazoyerRapaport}
J.~Mazoyer and I.~Rapaport.
\newblock {Inducing an order on cellular automata by a grouping operation}.
\newblock In {\em Proceedings of STACS'98, in Lecture Notes in Computer
  Science}, volume 1373, pages 116--127. Springer, 1998.

\bibitem{MoritaIntrinsicUniv1D}
K.~Morita.
\newblock {Reversible simulation of one-dimensional irreversible cellular
  automata}.
\newblock {\em Theoretical Computer Science}, 148(1):157--163, 1995.

\bibitem{MoritaCompUniv1D}
K.~Morita and M.~Harao.
\newblock {Computation universality of one-dimensional reversible (injective)
  cellular automata}.
\newblock {\em IEICE Trans. Inf. \& Syst., E}, 72:758--762, 1989.

\bibitem{MoritaCompUniv2D}
K.~Morita and S.~Ueno.
\newblock {Computation-universal models of two-dimensional 16-state reversible
  cellular automata}.
\newblock {\em IEICE Trans. Inf. \& Syst., E}, 75:141--147, 1992.

\bibitem{NagajWocjan}
D.~Nagaj and P.~Wocjan.
\newblock {Hamiltonian Quantum Cellular Automata in 1D}.
\newblock ArXiv preprint: arXiv:0802.0886, 2008.

\bibitem{OllingerJAC}
N.~Ollinger.
\newblock {Universalities in cellular automata a (short) survey.}
\newblock In B.~Durand, editor, {\em First Symposium on Cellular Automata
  ``Journ{\'e}es Automates Cellulaires'' (JAC 2008), Uz{\`e}s, France, April
  21-25, 2008. Proceedings}, pages 102--118. MCCME Publishing House, Moscow,
  2008.

\bibitem{PazZurek}
J.~P. Paz and W.~H. Zurek.
\newblock {Environment-induced decoherence and the transition from quantum to
  classical}.
\newblock {\em Lecture Notes in Physics}, pages 77--140, 2002.

\bibitem{PerezCheung}
C.A. P{\'e}rez-Delgado and D.~Cheung.
\newblock {Local unreversible cellular automaton ableitary quantum cellular
  automata}.
\newblock {\em Physical Review A}, 76(3):32320, 2007.

\bibitem{Raussendorf}
R.~Raussendorf.
\newblock {Quantum cellular automaton for universal quantum computation}.
\newblock {\em Physical Review A}, 72(2):22301, 2005.

\bibitem{SchumacherWerner}
B.~Schumacher and R.~Werner.
\newblock {Reversible quantum cellular automata.}
\newblock ArXiv pre-print quant-ph/0405174, 2004.

\bibitem{ShepherdFranz}
D.~J. Shepherd, T.~Franz, and R.~F. Werner.
\newblock {A universally programmable quantum cellular automata}.
\newblock {\em Phys. Rev. Lett.}, 97(020502), 2006.

\bibitem{Theyssier}
G.~Theyssier.
\newblock {Captive cellular automata}.
\newblock In {\em Proceedings of MFCS 2004, in Lecture Notes in Computer
  Science}, volume 3153, pages 427--438. Springer, 2004.

\bibitem{ToffoliConstruction}
T.~Toffoli.
\newblock {Computation and construction universality of reversible cellular
  automata}.
\newblock {\em J. of Computer and System Sciences}, 15(2), 1977.

\bibitem{VanDam}
W.~Van~Dam.
\newblock {Quantum cellular automata}.
\newblock Masters thesis, University of Nijmegen, The Netherlands, 1996.

\bibitem{VollbrechtCirac}
K.~G.~H. Vollbrecht and J.~I. Cirac.
\newblock {Reversible universal quantum computation within
  translation-invariant systems}.
\newblock {\em New J. Phys Rev A}, 73:012324, 2004.

\bibitem{Neumann}
J.~von Neumann.
\newblock {\em {Theory of Self-Reproducing Automata}}.
\newblock University of Illinois Press, Champaign, IL, USA, 1966.

\bibitem{Watrous}
J.~Watrous.
\newblock {On one-dimensional quantum cellular automata}.
\newblock {\em Complex Systems}, 5(1):19--–30, 1991.

\bibitem{WatrousFOCS}
J.~Watrous.
\newblock On one-dimensional quantum cellular automata.
\newblock {\em Foundations of Computer Science, Annual IEEE Symposium on},
  528537:528--537, 1995.

\end{thebibliography}
